\documentclass[11pt]{article}
\usepackage{erice,epsfig}

\def\GG{\gamma \gamma}

\def\be{\begin{equation}}
\def\ee{\end{equation}}
\def\bea{\begin{eqnarray}}
\def\eea{\end{eqnarray}}

\begin{document}
\vspace*{4cm}
\title{PHYSICS OPPORTUNITIES IN  ULTRAPERIPHERAL 
HEAVY ION  COLLISIONS AT LHC}

\author{ G.Baur }

\address{Institut f\"ur Kernphysik\\
Forschungszentrum Juelich,Germany}

\maketitle\abstracts{
Due to coherence, there are strong electromagnetic
fields of short duration in very peripheral 
heavy ion collisions. They give rise to
photon-photon and photon-nucleus collisions with high flux up to an
invariant mass region hitherto unexplored experimentally.
Photon-photon and photon-hadron
 physics at
various invariant mass scales are  discussed. 
Due to the very strong electromagnetic fields there
are interesting  many-photon 
exchange processes like  double giant 
dipole resonance excitation in nuclei, multiple $e^+e^-$
pair production and vector meson pair production.
Maximum equivalent photon energies
in the lab-system(collider frame) are typically of the order of 3 GeV for RHIC 
and 100 GeV for LHC.  Vector meson production(coherent as well as 
incoherent) and 
photon-gluon fusion leading to 
heavy quark jets are processes of great physics interest.
The high photon-photon luminosities in 
very peripheral collisions  at the LHC
will  allow to study interesting 
physics channels in the invariant mass 
region up to about 100 GeV. It is hoped that this 
will bridge a gap to the physics at the future 
photon colliders.
In pp collisions this 
maximum energy is even higher, and their is also the  
experimental possibility to tag on these photons.  
}

\section{Introduction}
In ultraperipheral (UPC) heavy ion collisions
nuclei  interact with each other through their
electromagnetic fields. In these collisions 
with impact parameter b larger than the sum of the nuclear radii
a new range of physical parameters enters. Electromagnetic fields 
are very strong and of short duration. The interaction 
can be very well described by the equivalent photon method.
Fluxes are very high and extend up to energies hitherto 
unexplored experimentally. 
A review article for Physics Reports 
on coherent $\gamma \gamma$ and $ \gamma$ A 
interactions in relativistic  
ion colliders  has just been completed \cite{BHTSK}.

It is the aim of this talk to discuss the physics 
of UPC. Emphasis is put on the LHC and RHIC energy regions.
Detectors at RHIC as well as at LHC are built with the aim to
study central heavy ion  or pp collisions.
The question of how to detect such UPC 
events at the colliders is not dealt with here.
An important  question is the  triggering on UPC
events. The STAR group at RHIC led by Spencer Klein
has shown how to 
trigger on these events and how to study them experimentally.
I refer to the discussion  in these proceedings
\cite{pablo,joakim}.

Since the electromagnetic fields are very strong,
many photons can be exchanged in these collisions.
 In section~\ref{sec:multi} processes where more than one photon 
is emitted from a nucleus are discussed. There are a few
important and interesting processes of this type. Generally, the 
one photon exchange mechanism   is sufficient for inelastic 
processes.

A very useful view on the electromagnetic processes is the parton
picture with the photons as the partons. 
In this picture the scattering is 
described as an incoherent superposition of the scattering of the various 
constituents. For example, nuclei consist of nucleons which in turn consist 
of quarks and gluons, photons consist of lepton pairs, electrons consist of
photons, etc.. Relativistic nuclei have photons as an important constituent. 
This is due to the coherent action of all the charges in the nucleus: for 
these conditions the wavelength of the photon is larger than the size of the
nucleus, therefore it does not resolve the individual nucleons but sees the 
coherent action of all of them. 

\begin{figure}
\begin{center}
\psfig{figure=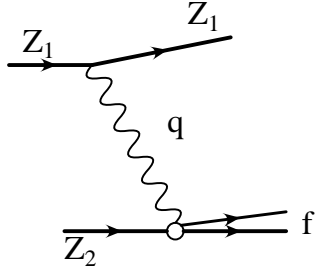,height=1.5in}
\end{center}
\caption{One photon exchange in inelastic ultraperipheral 
heavy ion collision. The exchanged photon is almost real,
i.e. $Q^2=-q^2 \sim 0$. 
\label{fig:fig1}}
\end{figure}

The coherence condition limits the virtuality $Q^2=-q^2$ of the photon to 
very low values
\begin{equation}
Q^2 < 1/R^2, 
\label{eq:cond_coherent}
\end{equation}
where the radius of a nucleus is
approximately $R=1.2\mbox{ fm}\ A^{1/3}$ with $A$ the nucleon
number. This is due to the rapid decrease of the 
nuclear electromagnetic form factor for high $Q^2$ values.
For most purposes these photons can therefore be considered as 
real (``quasireal''),see Fig.1.
From the kinematics of the process one has a photon four-momentum of 
$q_\mu=(\omega,\vec q_\perp,q_3=\omega/v)$, where $\omega$ and $q_\perp$ 
are energy and transverse momentum of the quasireal photon in a given
frame, where the projectile moves with velocity $v$. This leads to an 
invariant four-momentum transfer of 
\begin{equation}
Q^2=\frac{\omega^2}{v^2\gamma^2}+q_\perp^2,
\end{equation}
where the Lorentz factor is $\gamma=E/m=1/\sqrt{1-v^2}$.
The condition Eq.~(\ref{eq:cond_coherent}) limits the maximum energy of the 
quasireal photon to 
\begin{equation}
\omega<\omega_{max} \approx \frac{\gamma}{R},
\label{eq_wmax}
\end{equation}
and the perpendicular component of its momentum to
\begin{equation}
q_\perp <  \frac{1}{R}.
\label{pt_max}
\end{equation}
At LHC energies ($\gamma=3000$) 
this means a maximum photon energy of about 
100~GeV in the laboratory system, at RHIC 
($\gamma=100$) this number is 
about 3~GeV.
In section~\ref{sec:ga} $ \gamma-$A interactions 
are discussed. 
 $\gamma \gamma$ 
processes are a (small) subset of such processes. They are
very important in general and they allow to study 
very interesting physics in a clean way. This is 
discussed in section~\ref{sec:gg}. 
Conclusions and an outlook are given in  section~\ref{sec:co}.
    
\section{Strong Field Effects, Exchange of Many Photons}\label{sec:multi}

The Sommerfeld parameter 
\begin{equation}
\eta=\frac{Z_1 Z_2 e^2}{\hbar v} 
\sim Z_1 Z_2 \alpha \sim Z_1 Z_2/137
\label{eq:eta}
\end{equation}
 characterizes the 
strength of the electromagnetic interaction between the ions
with charges $Z_1$ and $Z_2$ respectively, see Fig. 2a. This parameter
$\eta$ can be $\gg1$,
e.g. for Pb-Pb collisions we have $\eta=49 $. 

\begin{figure}
\begin{center}
\psfig{figure=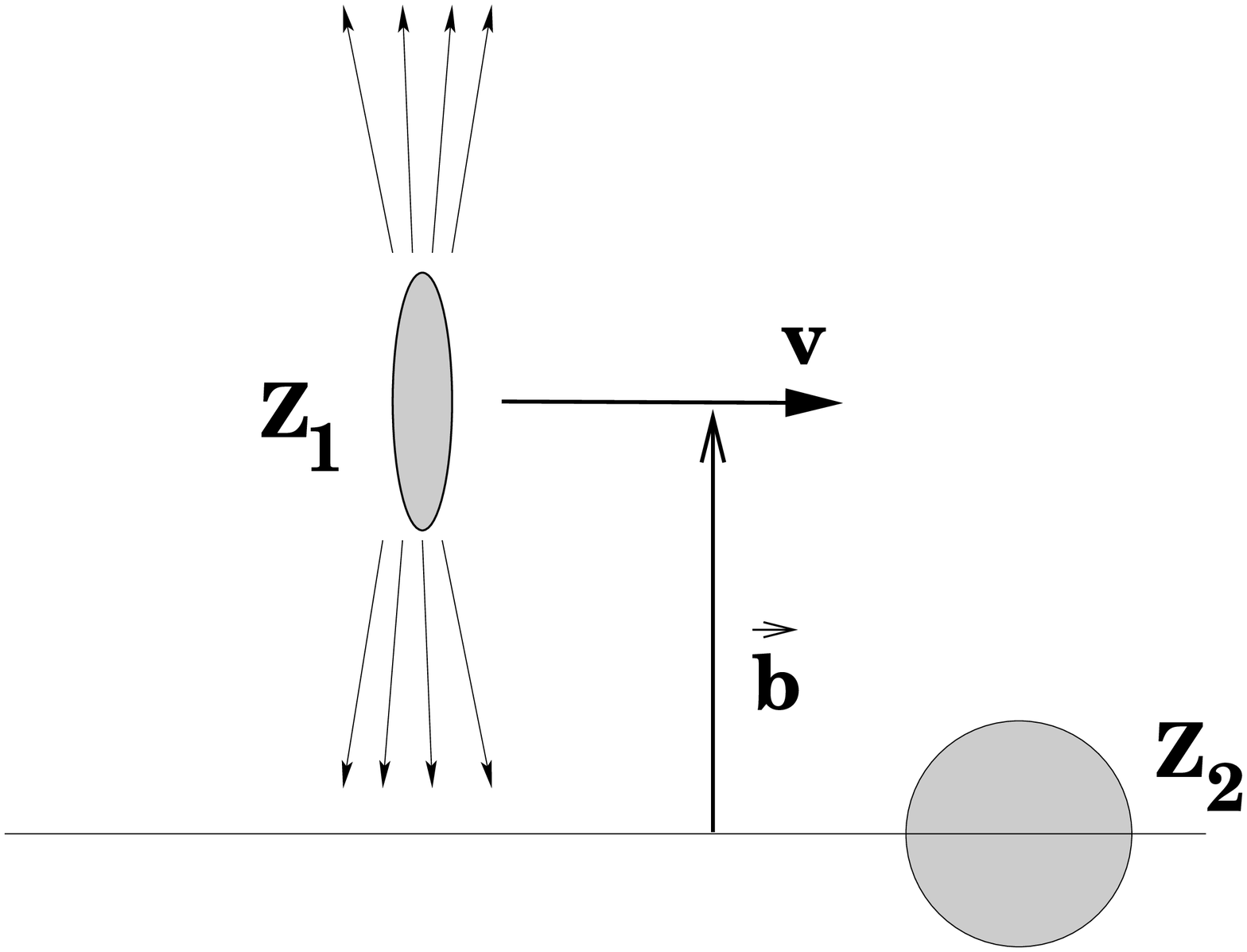,height=1.5in}
a)\hfil
\psfig{figure=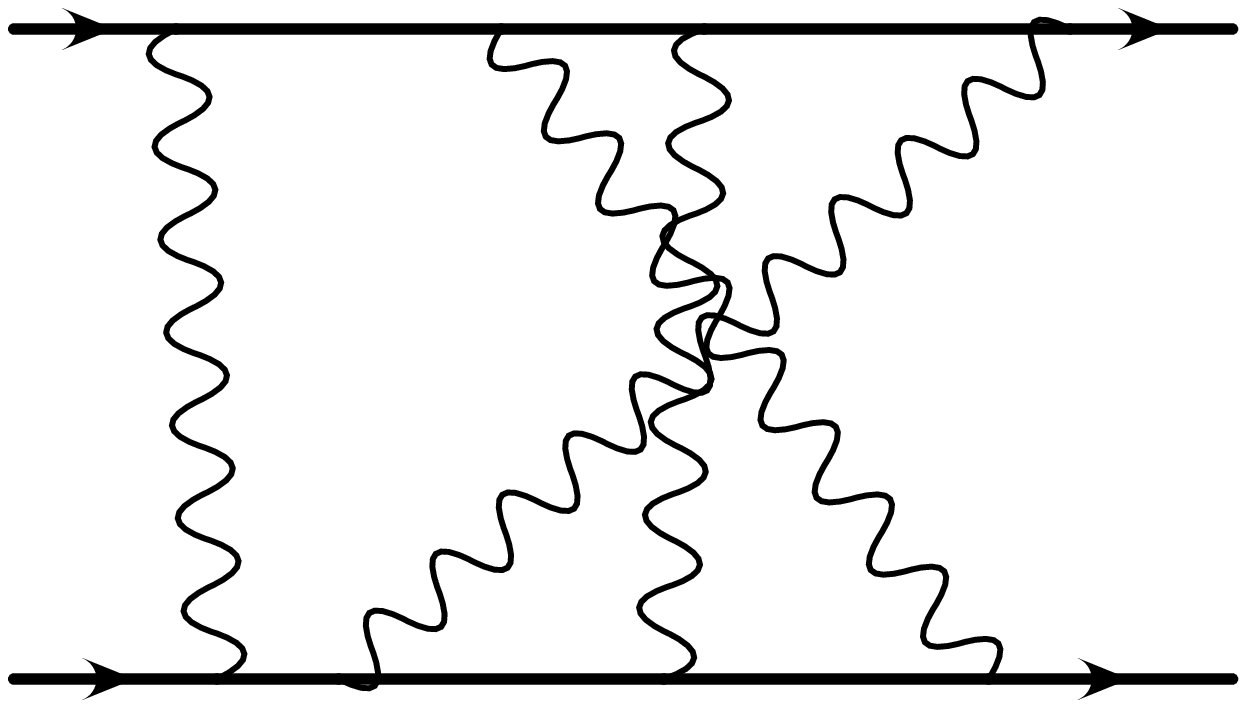,height=1.5in}
b)
\end{center}
\caption{a) Classical picture of heavy ion scattering.
This is also used by Fermi in his celebrated 
paper on the equivalent photon approximation
\protect\cite{Fermi}.
He considered the nonrelativistic case $v<<c$. The relativistic
case is discussed in a 
very pedagogical manner in  \protect\cite{jackson}. 
b) For $\eta \gg1$ many photons are exchanged in 
elastic heavy ion scattering. A typical graf is shown.  
\label{fig:fig2}}
\end{figure}
  
According to 
classical arguments given by N.Bohr elastic scattering 
can be considered as classical for $\eta \gg1$.
 For relativistic heavy ions the classical trajectory
is almost a straight line with an impact parameter b. 
The momentum transfer $\Delta p$ is essentially 
perpendicular to the 
beam direction and is given by 
\begin{equation}
\Delta p=\frac{2Z_1 Z_2e^2}{bv}=\frac{2\eta}{b} \hbar
\label{eq:deltap}
\end{equation}
This momentum transfer is built up from the exchange 
of many photons \cite{HaItz}, see Fig. 2b. Due to the large value 
of $\eta$ the integral over the impact parameter b
in the Glauber amplitude can be evaluated using the 
stationary phase (or saddle point) approximation.
This phase is given by 
\be
\phi=-qb+2\eta \ln(kb)
\ee
where k is the wave number of the projectile nucleus.
The condition $\phi'(b)=0$ leads to 
$b=\frac{2\eta}{q}$, i.e. $\Delta p=\hbar q$.
Thus the momentum transfer q(or scattering angle 
$\theta =\frac{q}{k}$
)
 is related to the classical impact parameter.
This is in contrast to e.g. p-p scattering
where $\eta=1/137 \ll1 $ and the Born approximation
(one photon exchange) is sufficient. For the heavy ions
the strong interactions 
for collisions with $b<R_{min}$, where 
$R_{min}$ is the sum of the nuclear radii,
lead to complete absorption
(black disk approximation). This disregards the 
diffuseness of the nucleus, but it is quite 
 a good approximation for many purposes. 
 In the case of strong Coulomb coupling($\eta\gg1$)
the elastic 
scattering cross section is given by a Fresnel
diffraction pattern(rather than a Fraunhofer one for $\eta<1$).
This is explained in Ref. \cite{Frahn72}, see also Sect 5.3.5
of Ref. \cite{NoeWei}. 

The electric charge of the relativistic ion gives rise
to an electromagnetic potential, the Lienard-Wiechert
potential $A_{\mu}(\vec{r},t)$. This potential interacts with 
a target current $j_{\mu}$. This target current
describes e.g. nuclear states, vector mesons or 
$e^+e^-$ pairs in the field of a nucleus. This defines
a time dependent interaction $V(t)$ as( see e.g. \cite{beba88})
\begin{equation}
V(t)=\int d^3r A_{\mu}(\vec{r},t) j_{\mu}(\vec{r})
\label{eq:vint} 
\end{equation} 
The dependence of $V(t)$ on the impact parameter 
is not shown explicitly.
Coupled equations for the excitation amplitudes
$a_n(t)$ for certain states $n$ can be set up.
The solution of these equations is  greatly 
facilitated if the sudden approximation can be applied,
see e.g. Ref.\cite{ba91}. 
This is the case if the collision
time is much smaller than the nuclear excitation
time. This condition is fulfilled in many interesting
cases and we assume now that the sudden approximation
can be used. This "frozen nucleus"-approximation 
is also used in  
Glauber theory. The relation between the 
semiclassical approach  and the (quantal) Glauber 
(or eikonal)approximation is explained in Ref.\cite{ba91}.
This is done for the non-relativistic as well 
as the relativistic case.
The excitation amplitude is given by
\begin{equation}
a_n(t \rightarrow \infty) =<n| \exp(iR)|0>
\label{eq:an}
\end{equation}
where $
R=-\int_{-\infty}^{+\infty} V(t)dt$ ( we put $\hbar$=1).
The operator R is a direct sum of operators in
the space of nuclear states, the space of the nucleus-
vector meson system, the nucleus-$e^+e^-$ system,etc.. 
We can expand the exponential in eq.~\ref{eq:an}.
Terms linear in R
give e.g. the excitation of nuclear states,like the 
collective giant dipole resonance(GDR), 
vector meson production  or $e^+e^-$pair production.
 Terms quadratic in R give e.g. contributions to double phonon
GDR-excitation , double vector meson production ,
two $e^+e^-$ pair production . It also describes 
e.g.  vector meson production and GDR excitation
in a single collision.
A contribution to the second order amplitude $a^{(2)}$ 
is e.g. 
\begin{equation}
a^{(2)}=-<GDR|R|0><\rho^0|R|0>
\label{eq:a2}
\end{equation}
where $|0>$ denotes the ground state of the nucleus.
The factor $1/2!$ in the expansion of $exp(iR)$, see eq.~\ref{eq:an},
is compensated by the two possibilities
in the time ordering of the GDR excitation and 
the vector meson production respectively, see Fig. 3.

\begin{figure}
\begin{center}
\psfig{figure=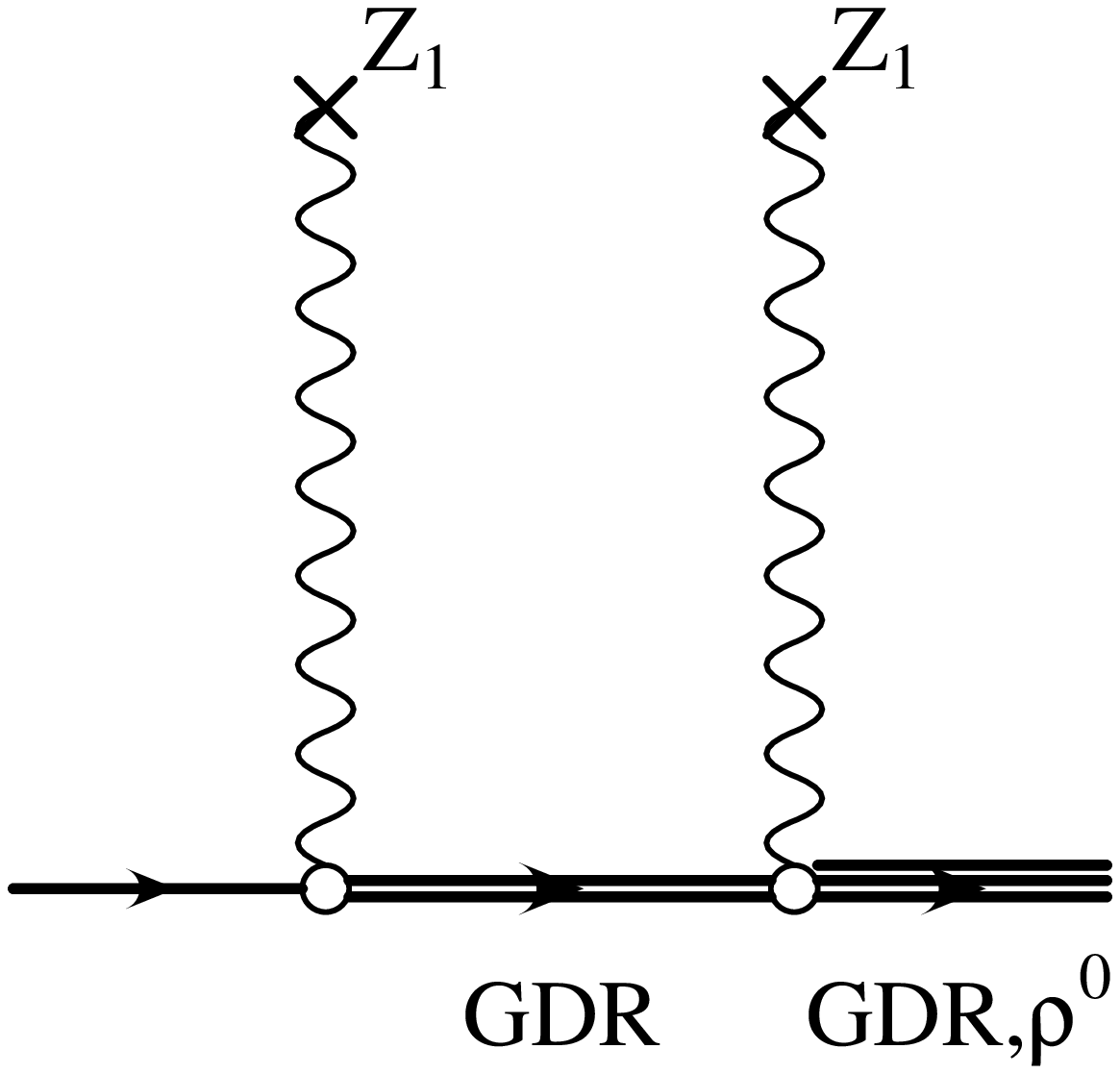,height=1.5in}
a)\hfil
\psfig{figure=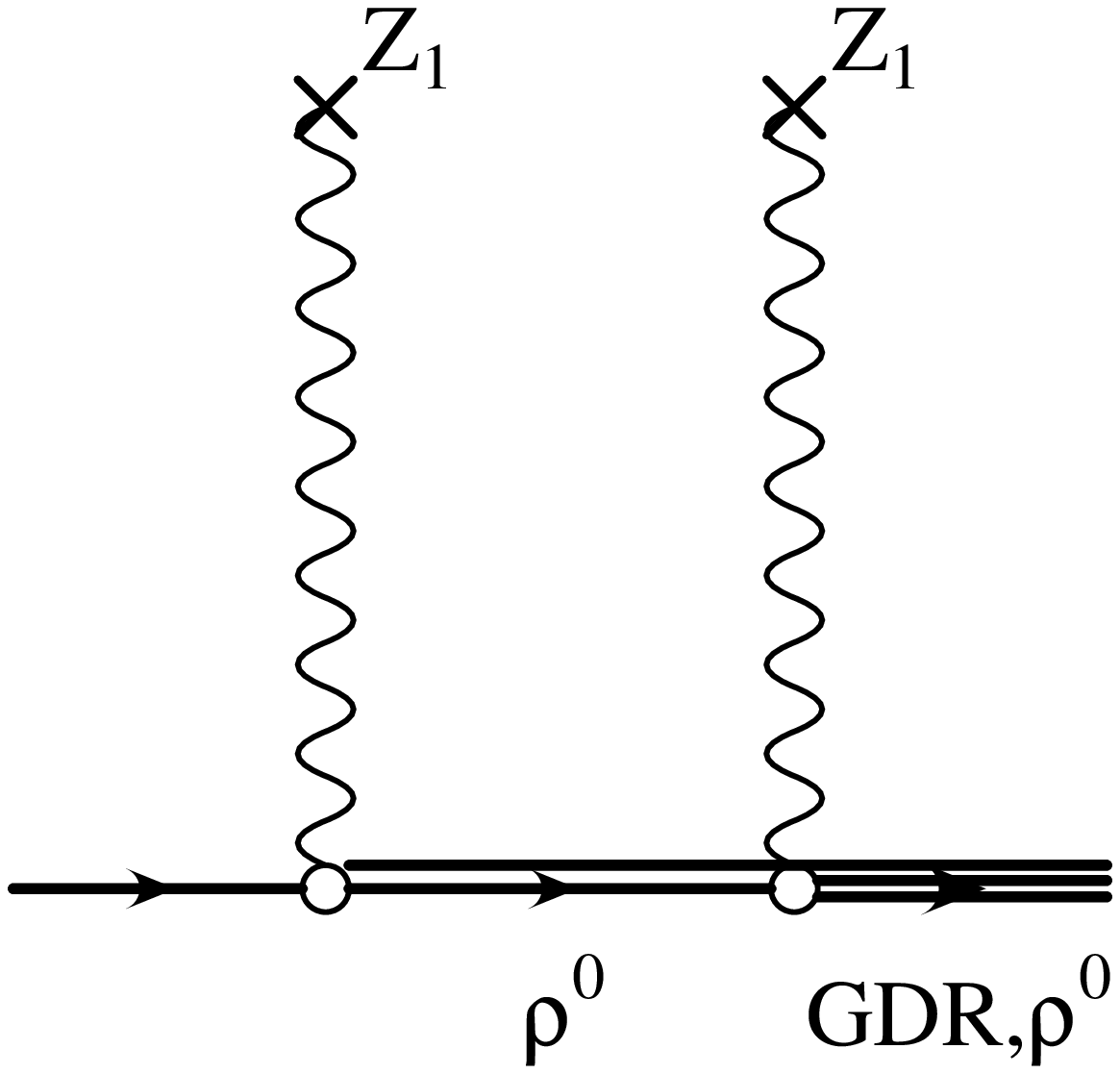,height=1.5in}
b)
\end{center}
\caption{ Grafs contributing to the simultaneous 
excitation of the GDR and $\rho ^0$ production.
\label{fig:fig3}}
\end{figure}

 For three independent
processes, say GDR-excitation, vector meson- and 
$e^+e^-$ pair production , there are 6 different 
time orderings, which compensate  the $\frac{1}{3!}$
factor in the expansion of eq.~\ref{eq:an}, and so on.
In this formalism it is clearly seen that these
processes are independent and the elementary amplitudes 
factorize, as one would have intuitively expected.
This property is used e.g. in the experimental analysis 
of vector meson production with  simultaneous 
GDR excitation. The neutrons from the GDR decay 
serve as a trigger on UPC \cite{pablo,joakim}. 
The ion motion is not disturbed by the excitation 
process. The reason is that the kinetic energy of
the ion is much larger than the excitation energy.

\subsection{Exchange of Many Photons in 
Multi-Phonon Giant Resonance Excitation}\label{subsec:dgdr}
 An especially simple and important case is the 
excitation of a harmonic oscillator. In terms
of the corresponding creation and destruction operators 
$a^{\dagger}$ and $a$ the Hamiltonian of the system
is given by 
\begin{equation}
H=\hbar\omega(a^{\dagger}a+\frac{1}{2})
\label{eq:hamil}
\end{equation}
where $\omega$ denotes the energy of the oscillator.
We have the boson commutation
rule $[a,a^{\dagger}]=1$.
Only one mode is shown explicitly, in general one has to 
sum (integrate) over all the possible modes. The excitation operator 
is assumed to be linear in the destruction and creation 
operators 
\begin{equation}
R=-(ua+u^*a^{\dagger})
\label{eq:r}
\end{equation}
where u is a c-number which characterizes the excitation process
(the matrixelement of R between the ground state 
and the one-phonon-state).
This leads to the excitation of a 
so-called coherent state, see \cite{glauber}.
For the excitation of multiphonon states this is 
explicitly shown in \cite{bebapadova}. One has
\begin{equation}
a_n=<n|e^{-i(u*a^{\dagger}+ua)}|0>=\frac{(-iu^*)^n}{\sqrt(n!)}e^{-\frac{1}{2}u u^*}
\label{eq:sudden}
\end{equation}
where the operator identity
$e^{A+B}=e^A e^B e^{-\frac{1}{2}[A,B]}$ was used, which 
is valid for two operators A($=-iu^*a^{\dagger}$)  and 
B($=-iua$) for which the 
commutator is a c-number. 

Electromagnetic excitation of nuclear states, especially the 
collective giant multipole resonances was  discussed
at this workshop by Carlos Bertulani \cite{carlos}. 
The possibility to 
excite multiphonon GDR states is discussed in 
\cite{babeerice}, where also its main properties
like decay widths are discussed. The parameter which describes
the probability $\Phi$ of GDR excitation is \cite{babePLB}
\begin{equation}
\Phi=\frac{2\alpha^2Z_1^2N_2Z_2}{A_2m_N\omega b^2}
\label{eq:phi}
\end{equation}
where $m_N$ denotes the nucleon mass, the neutron-
proton-, and mass-number of the excited nucleus
are given by $N_2,Z_2,$ and $A_2$ respectively.
The excitation probability is inversely proportional 
to the energy $\omega (\sim 80MeVA^{-1/3})$ of the GDR state. Thus
soft modes are more easily excited, as one may have expected.
In this (rather accurate) estimate, it was assumed that 
the classical dipole sum rule (Thomas-Reiche Kuhn sum rule)
is exhausted to 100 percent. For the excitation of an N-phonon state,
 a Poisson distribution is obtained. For the heavy systems
$\Phi$ is of the order of $\frac{1}{2}$ for close collisions
($b \sim R_1+R_2$).

Quite similarly , double $\rho^0$ production was studied in Ref.
\cite{sk99}. In addition to the label m for the magnetic substates
of the GDR, one has a continuous label (the momenta) 
in the case of vector meson production. The probability 
to produce a vector meson in a close collision 
is of the order of one to three percent for the heavy systems.    

\subsection{Production of Multiple Electron-Positron 
Pairs}

In Ref. \cite{pra} it was shown that multiple $e^+e^-$-pairs
can be produced in relativistic heavy ion collisions.
In this work the  sudden (or Glauber) approximation and a
quasiboson approximation for $e^+e^-$ pairs was assumed.
This will be  well fulfilled in practice. Using a  
QED calculation (including Coulomb corrections in the 
Bethe-Maximon approach) for one pair production as an 
input, a Poisson distribution is obtained for multiple 
pair production. This is quite natural, since this problem is now 
reduced to the excitation of a harmonic oscillator
(the modes are labelled by the spins and 
momenta of the $e^+e^-$ pairs)
, see above. 

The characteristic dimensionless parameter for this problem is 
$\Xi= \frac{(Z_1 Z_2 \alpha^2)^2}{(mb)^2}$ where m is 
the electron mass and $b>1/m$.
For impact parameters b$\sim 1/m$ and heavy systems like
Pb-Pb or Au-Au the parameter $\Xi$ is of the order of 
unity.
 In a series of papers by K.Hencken
et al. (see \cite{he+-}) the impact parameter dependence 
of $e^+e^-$ pair production was studied numerically in lowest
order QED. Only recently an approximate analytical 
formula for the total pair production probability
in lowest 
order $P^{(1)}$was found. In an impact
parameter range of $1/m<b<\gamma/m$ it is given by \cite{lms}
\begin{equation}
P^{(1)}= \frac{28}{9\pi^2}\Xi(2\ln\gamma^2-3\ln(mb))\ln(mb)
\label{eq:p1}
\end{equation}    
The N pair-production probability decreases strongly with 
increasing impact parameter b 
(approximately like 
$\sim b^{-2N}$). Therefore the probabilities $P^{(1)}(b)$
should be known accurately for an impact parameter
 range of $0<b< $several $1/m$. 

Muon pair production is also of interest. In close collisions, 
the pair production probability is of the order of 1, as can
be seen from eq. 15. For a more exact evaluation, form 
factor effects would have to be taken into account. Due to 
the much smaller Compton wave length of the muon, total cross
sections are smaller than about a factor of $(m_e/m_{\mu})^2 
\sim 40 000$.   

\section{$\gamma$A interactions}\label{sec:ga}

It seems appropriate at this point to recall 
various interesting 
results which have been obtained 
in the past decade in  UPC experiments at lower energies \cite{beba88}.
References to these experiments can be found e.g. in \cite{bhttwerice}. 
At beam energies in the GeV region, the equivalent photon 
spectrum extends up to $\omega_{max}=\frac{197MeVfm}{R_{min}}\gamma
\sim 10-20 MeV$. With such a soft spectrum of photons up to the 
GDR region, many interesting topis in nuclear structure physics
and astrophysics could be investigated. The excitation of 
the DGDR was already briefly 
discussed above in subsection~\ref{subsec:dgdr}:
strong (quasireal)photon sources are essential for such investigations. 
Another application is the study of radiative capture processes:
they are related to photodissociation via time reversal \cite
{bhttwerice,bbr}. E.g. the astrophysical S-factor for the 
$^7Be(p,\gamma)^8$B reaction could be studied with the 
electromagnetic dissociation of  $^8B$ rare isotope 
beams at RIKEN(Japan),GSI(Darmstadt) 
and Michigan State University.
 This is of great interest for the solar neutrino problem:
most of the neutrinos detected in the Kamiokande or 
SNO detectors originate from the weak decay of $^8B$.
This nucleus is solely produced in the $^7 Be(p,\gamma)$ capture 
reaction,
thus this astrophysical S-factor directly determines the 
(high energy) neutrino flux from the sun.

While GDR excitation and subsequent 
particle(mainly neutron) decay  is a source of beam loss,
a means to trigger on UPC 
as well as a tool  for a luminosity monitor \cite{sebastian}
("yesterday's sensation is today's calibration"), the 
physics interest concentrates on the higher equivalent 
photon energies available at RHIC(up to about 600GeV) and LHC
(up to about 600 TeV), as viewed from the nucleus rest frame.
The physics is quite similar to what is done at HERA,
with essentially two  differences: now the photons are
only quasireal($Q^2\sim 0$) and the hadron is now 
 a nucleus instead of a proton.
There is (diffractive) coherent and incoherent
vector meson production with many interesting applications.
Indeed, relativistic heavy ion colliders are 
vector meson factories \cite{capri,spencer}. This opens 
the way for interesting experimental studies, like the interference
effect in exclusive vector meson production \cite{spenprl}.
As explained in this refererence, this provides also an example
of the Einstein-Podolsky-Rosen paradoxon. Double $\rho ^0$
production \cite{sk99}was already mentioned above and the 
prospects for vector meson spectroscopy were discussed 
in \cite{capri}. Certainly, the forthcoming experimental 
results will spur further theoretical developments. 
   
Another interesting topic is photon-gluon fusion into 
$q\overline{q}$. The two jets can be identified and the kinematics 
resconstructed. These processes were first studied in Ref.
\cite{soff} and Ref. \cite{baron}. Quite recently, it was suggested to 
investigate the colour glass condensate with this method
\cite{gel1,gel2}.
In this way, the low-x gluon distribution  can be directly 
studied 
experimentally . 
Top quark production will also be possible  \cite{topram}. 

The points above are most important also for the future
experiments at RHIC and LHC. I only refer 
here to  \cite{capri} and \cite{BHTSK}.
Further  discussions and calculations are also reported 
at this workshop in  \cite{kai}.

\section{$\gamma \gamma$ physics }
\label{sec:gg}

At LHC ultraperipheral collisions
provide equivalent photons with energies up to 100 GeV in the 
c.m. system.  
$\gamma \gamma $-collisions have been  
investigated at LEP in the invariant mass region up to 
185 GeV. The effective $\gamma \gamma $
luminosity for medium mass heavy ions exceeds the one 
achieved at LEP, see e.g. Fig.6 of the recent review
\cite{BHTSK}. A review of 
$\gamma \gamma$ physics is also given there, so I will only mention
a few considerations here, see also the talk of Kai Hencken
\cite{kai}.
These very high effective $\gamma \gamma$ luminosities 
will allow QCD studies like the study of $\gamma \gamma$
widths of $C=+1$ (heavy) mesons and the total cross 
section for $\gamma \gamma \rightarrow $ hadrons,see Fig.4.  
Also the study of  new physics will be possible.

\begin{figure}
\begin{center}
\psfig{figure=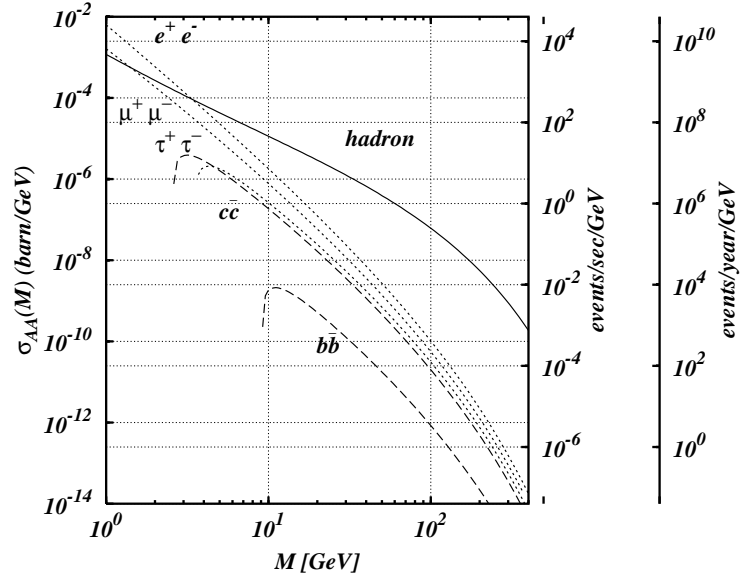,height=3in}
\end{center}
\caption{Cross section per GeV for fermion pair production 
at the LHC for Ca-Ca collisions. A luminosity of 
$L_{CaCa}=4x10^{30}cm^{-2}s^{-1}$ was assumed. The process $\gamma \gamma
\rightarrow hadrons $ is also shown. 
A "(heavy ion) year" corresponds to $10^6$ s.
For further details 
see \protect\cite{BHTSK}
\label{fig:fig4}}
\end{figure}

One can speculate about new particles with strong coupling to the
$\GG$-channel. Large $\Gamma_{\GG}$-widths will directly 
lead to large $\GG$ production cross-sections. 
The two-photon width of quarkonia is proportional to the wave function 
squared in the center of the system. Thus we 
can expect, that if a system is very tighty bound it should have 
a large two-photon width due to the factor $|\Psi(0)|^2$
which is large in these cases.
Examples for such tightly bound systems were  discussed 
 in the eighties e.g. in Refs. \cite{Renard83,BaurFF84}.
Composite scalar bosons at $W_{\GG}\approx 50$~GeV are expected to have 
$\GG$-widths of several MeV \cite{Renard83,BaurFF84}. 
The search for such kind of resonances in the $\GG$-production channel 
will be possible at LHC.
In the TESLA Report part3 p.110 (Ref. \cite{Tesla})
 the reader can find some 
remarks about the "agnostic" approach to compositeness. 
From eq. 68 and figure 14 of Ref. \cite{BHTSK}
one can directly obtain a value for the production cross-sections
of such states and the corresponding rates in the various collider
modes.
Due
to the high flux of equivalent photons such searches 
seem worth-while, 
and a possible discovery would be quite interesting.

Certainly one will have to wait for the future  $\GG$ colliders
in order to study $\GG$- 
 physics in the region of several 100 GeV. However,
it may be possible to have a glimpse into this region with the 
(ultra)peripheral collisions at LHC which will 
be working (taking data) in a few  years from now.  

With their high luminosities pp collisions 
are also a very interesting source of equivalent photons. 
Their energy spectrum extends up to very 
high photon energies beyond 200 GeV.
In Refs. \cite{Zerwas,Drees} $\GG$-processes at pp colliders
are studied. It is observed there that non-strongly 
interacting supersymmetric particles(sleptons,
charginos, neutralinos and charged Higgs bosons) are 
difficult to detect in hadronic collisions at LHC.
The possibility of producing such particles in $\GG$
interactions is examined. Clean events can be generated 
which should should compensate for the small production number.   
This is even more important since 
there is a new experimental approach 
to tagging in pp coillisions by measuring very 
forward proton scattering \cite{piotr,pioeri}. 
Particularly exciting is the possibility to detect 
Higgs boson production via the $\GG$ fusion \cite{piotr}.

In \cite{GinzburgS98}
it was proposed to search for heavy magnetic monopoles in $\GG$
collisions 
at hadron colliders like the Tevatron and LHC. The idea is that 
        photon-photon 
scattering below the monopole production threshold 
is enhanced due to the strong coupling of magnetic monopoles to 
photons. The magnetic coupling strength $g$ is given by
$ g= \frac{2\pi n}{e} $ where $n=\pm1, \pm2,...$. Since 
$\frac{e^2}{4 \pi}=1/137$ the magnetic coupling strength is
indeed quite large. In this reference differential cross sections
for $\GG$ scattering via the monopole loop are calculated for
energies below the monopole production threshold. The result depends strongly
on the assumed value of the spin of the monopole. With this elementary
cross section as an input, the cross section for the process 
$ pp\rightarrow \GG X$ is calculated. Elastic(i.e.$ X=pp$ )
and inelastic contributions are taken into account. The signature of 
such a process is the production of two photons where the transverse
momentum of the pair is much smaller than the transverse momentum of
the individual photons. 
At the Tevatron such a search was performed. They looked at a pair of photons
with high transverse energies. No excess of events above background was found 
\cite{Abbott98}.
Thus a lower limit on the mass of the magnetic monopole could 
be given. A mass of 610, 870, or 1580 $GeV/c^2$ was obtained, for
the assumed values of a monopole spin of $0,~1/2,$ or $1$ respectively.
For further discussions see also \cite{mono}.
Such kind of searches could also be performed at the heavy ion colliders:
the flux is enhanced by a factor of  $Z^4$, on the 
other hand, the maximum equivalent photon energy is  lower.
Some exploratory theoretical calculations would be useful. 

\section{Conclusions. Outlook}\label{sec:co} 
Ultraperipheral heavy ion collisions provide  a strong 
source of equivalent photons up to very high energies.
This offers the unique possibility to study 
photon-hadron(nucleus) and photon-photon processes
in hitherto inaccessible regions.
 Some of these comparatively silent (as compared to 
the violent central collisions) events
are very interesting.
With the forthcoming 
experimental results from RHIC the field 
of ultraperipheral collisions is rapidly expanding.
These results will also be helpful for the planning
of UPC experiments at LHC. 
This kind of physics
 is of great interest and potential.

\section*{Acknowledgments}
I wish to thank Sebastian White and Bill Marciano for  
inviting me  to come to Erice to 
a very  stimulating
 workshop
in a most pleasant atmosphere.

\end{document}